\documentclass[12pt, referee]{aastex}
\usepackage[dvips]{color}
\newcommand{\re}{\textcolor[rgb]{0,0,0}}

\shorttitle{Interstellar argonium}
\shortauthors{Neufeld and Wolfire}
\def\gtsima{$\;\buildrel > \over \sim \;$}
\def\simgt{\lower.5ex \hbox{\gtsima}}
\def\ltsima{$\;\buildrel < \over \sim \;$}
\def\simlt{\lower.5ex \hbox{\ltsima}}

\begin{document}

   \title{The chemistry of interstellar argonium and other probes of the molecular fraction in diffuse clouds}
\author{David A. Neufeld\altaffilmark{1} and Mark G. Wolfire\altaffilmark{2}}
\altaffiltext{1}{Department of Physics and Astronomy, Johns Hopkins University,
3400 North Charles Street, Baltimore, MD 21218; neufeld@pha.jhu.edu}
\altaffiltext{2}{Department of Astronomy, University of Maryland, College Park, MD 20742; mwolfire@astro.umd.edu}

  \begin{abstract}
{We present a general parameter study, in which the abundance of interstellar argonium (ArH$^+$) is predicted using a model for the physics and chemistry of diffuse interstellar gas clouds. Results have been obtained as a function of UV radiation field, cosmic-ray ionization rate, and cloud extinction. No single set of cloud parameters provides an acceptable fit to the typical ArH$^+$, OH$^+$ and $\rm H_2O^+$ abundances observed in diffuse clouds within the Galactic disk.  Instead, the observed abundances suggest that ArH$^+$ resides primarily in a separate population of small clouds of total visual extinction of at most $0.02$~mag \re{per cloud}, within which the column-averaged molecular fraction is in the range $\re{10^{-5} - 10^{-2}}$, while OH$^+$ and $\rm H_2O^+$ reside primarily in somewhat larger clouds with a column-averaged molecular fraction $\sim 0.2$.  
This analysis confirms our previous suggestion (S14) that the argonium molecular ion is a unique tracer of almost purely {\it atomic} gas.
}
\end{abstract}
%

\section{Introduction}

One of the more unexpected results obtained from the Heterodyne Instrument for the Infrared (HIFI) on board the {\it Herschel Space Observatory} has been the discovery of interstellar argonium ions (ArH$^+$), the first known astrophysical molecule to contain a noble gas atom.Detected early in the {\it Herschel} mission as a ubiquitous interstellar absorption feature of unknown origin at 617.5~GHz (M\"uller et al.\ 2013), the $J=1-0$ transition of ArH$^+$ was finally identified by Barlow et al.\ (2013), following its detection, in emission, from the Crab nebula, along with a second transition (the $J=2-1$ transition) at twice the frequency.  While the {\it Herschel} detections of argonium were all obtained within the Galaxy,  
an extragalactic detection has been reported recently (M\"uller et al.\ 2015), within the $z=0.89$ galaxy along the sight-line to PKS 1830-211.

As discussed by Schilke et al.\ (2014; hereafter S14), several ``accidents" of chemistry conspire to permit the presence of interstellar argonium at observable abundances: ArH$^+$ has an unusually slow rate of dissociative recombination with electrons (Mitchell et al.\ 2005) and an unusually small photodissociation cross-section for radiation longward of the Lyman limit (Alekseyev et al.\ 2007).  Moreover, while other abundant noble gas 
cations (He$^+$ and Ne$^+$) undergo an exothermic dissociative ionization reaction when they react with with H$_2$, e.g. 
$$\rm Ne^+ + H_2 \rightarrow Ne + H + H^+, \eqno(R1)$$ 
the analogous reaction is endothermic in the case of Ar$^+$ and thus the alternative argonium-producing abstraction reaction is favored:
$$\rm Ar^+ + H_2 \rightarrow ArH^+ + H. \eqno(R2)$$  These unusual features of the relevant chemistry also suggested that argonium ions can serve as a unique tracer of gas in which the H$_2$ fraction is extremely small; indeed, the theoretical predictions of our diffuse cloud model (S14) indicated that the argonium abundance peaks very close to cloud surfaces where the molecular hydrogen fraction, $f({\rm H}_2)=2n({\rm H}_2)/n_{\rm H}$ is only 
$10^{-4}$ to $10^{-3}$.

In this paper, we present a general parameter study of argonium in diffuse clouds that extends the results we presented previously in S14.  In the present study, we computed the argonium column densities expected as a function of several relevant parameters: the gas density, the UV radiation field, the cosmic-ray ionization rate (CRIR), and the visual extinction across the cloud.
Our model is described in \S 2, and the results are presented in \S 3, together with analogous results for OH$^+$ and H$_2$O$^+$.  In \S 4, we discuss the results and describe how observations of all three molecular ions can be used to constrain the molecular fraction and cosmic ray ionization rate in the diffuse interstellar medium.

\section{Model}

\subsection{Diffuse cloud model}

Our thermal and chemical model for diffuse molecular clouds is based upon the model described by Hollenbach et al. (2012; hereafter H12), with the chemistry of argon-containing species added as described by S14. 
We adopted a two-sided slab model, 
of total assumed optical depth $\rm A_V(tot)$, in which the
radiation field in free space, $\chi_{\rm UV}$, is isotropically incident with a value of
$\chi_{\rm UV}$/2 on each side. 
Here, the UV interstellar radiation field (ISRF), $\chi_{\rm UV}$, is normalized with respect to the average field given by Draine (1978), and assumed to have the same spectral shape as that field.
The attenuation of the isotropic field was
calculated as described in Wolfire et al.\ (2010),
and the equilibrium gas temperature and steady-state chemical abundances 
were calculated as a function of
depth into the cloud.  We adopted gas-phase abundances, relative to H nuclei, of
$1.6 \times 10^{-4}$ for carbon nuclei (Sofia et al. 2004\re{; 
Gerin et al.\ 2015}) and
$3.9 \times 10^{-4}$ for oxygen nuclei (Cartledge et al. 2004).  

As described below, several collisional
excitation rates, heating rates, and chemical rates have been modified
relative to the H12 model.
For collisional excitation of C$^+$ by H$_2$ and by H, we adopted rate coefficients from
Wiesenfeld \& Goldsmith (2014) and from Barinovs et al.\ (2005), respectively.
For collisional excitation of atomic O by H, we adopted the rates from
Abrahamsson et al.\ (2007).  We have corrected a minor error that resulted in far-UV radiation being double-counted in the computation of the heating rate due to photodetachment of PAH$^-$.
Changes to the chemistry include the addition of a $^{13}$C chemical network, following
Burton et al. (2015), and the inclusion of cosmic-ray induced 
photoionization of C, using the
UMIST rate (McElroy et al.\ 2013) weighted by the molecular H$_2$ fraction.
For the reaction of H$_3^+$ with O, we adopted the rate coefficients 
of de Ruette et al.\ (2016),
and considered the product channels to OH$^+$ + H$_2$ and to $\rm H_2O^+$ + H.
We took the binding energies of O and OH on grain
surfaces, $E_b/k$, to
be 1800 K, and 4800 K respectively (He et al. 2015).
We adopted an HCl$^+$ dissociative recombination rate from Novotny et al.
2013). We also included the photodissociation of HCl$^+$ and
CF$^+$. For HCl$^+$ we adopted an unshielded rate of $9.9 \times 10^{-11} \chi_{\rm UV}$
(http://home.strw.leidenuniv.nl/$\sim$ewine/photo/),
with a depth-dependence proportional to ${\rm E}_2(3.5 A_{\rm V})$, where $\rm E_2$ is an exponential integral, the same as that adopted for OH$^+$. For
\re{CF$^+$}, we assumed an unshielded
rate of $2.2 \times 10^{-10} \chi_{\rm UV}$, 
with a depth-dependence the same as that for HCl, viz.\ ${\rm E}_2(2.1 A_{\rm V}).$

\subsection{Chemistry of argon-bearing species}

In the case of argon-containing species, we adopt the same network of chemical reactions that we presented in S14, with the adopted rates exactly as listed in Table 2 of 
that paper.\footnote{\re{We note here an error in the products given for the second reaction listed in that table 
(reaction of Ar with H$_2^+$): the products should have been given 
as ArH$^+$ and H.}} A complete discussion will not be given here, but the broad features may be described as follows.  The reaction network is initiated by the ionization of argon atoms by cosmic rays, and the resultant ions react rapidly with H$_2$ to form ArH$^+$ (reaction R2 in \S 1 above).When $f({\rm H}_2) \simlt 10^{-5}$, recombination of Ar$^+$ with electrons and PAH anions competes with ArH$^+$ formation, but otherwise every ionization of Ar is followed by the formation of argonium.  Argonium, which has a proton affinity of only 369 kJ~mol$^{-1}$, is destroyed by proton transfer to neutral species, including O, CO and,  
most importantly, H$_2$.  Photodissociation is potentially important as an additional destruction mechanism in regions of enhanced UV field.

\subsection{Standard model grid}

In S14, we obtained predictions for the ArH$^+$ abundance and column density for a single value of the gas density, UV radiation field, and CRIR.  In the present study, we
have constructed a large grid of diffuse cloud models, and thereby investigated the dependence of the ArH$^+$ column density on relevant physical parameters.  In principle, there are key four parameters in the model that affect the ArH$^+$ column density: 
\vskip 0.1 true in
\noindent (1) the ISRF, $\chi_{\rm UV}$,
\vskip 0.1 true in
\noindent (2) the CRIR, $\zeta_p({\rm H}),$ expressed as the primary ionization rate for atomic hydrogen.  The total ionization rate for atomic hydrogen is roughly a factor of 1.5 as large as $\zeta_p({\rm H})$ , and the total ionization rate for H$_2$ is roughly $2.3\, \zeta_p({\rm H})$ (Glassgold \& Langer 1973), the exact factors depending upon the fractional ionization and treated in our model using results given by (Dalgarno et al.\ 1999), 
\vskip 0.1 true in
\noindent (3) the total visual extinction through the cloud, $A_{\rm V}({\rm tot})$; and
\vskip 0.1 true in
\noindent (4) the gas density, $n_{\rm H}$, given as the volume density of H nuclei and assumed constant within the cloud; 
\vskip 0.1 true in
However, provided the gas densities are small enough that the total cooling rate is unaffected by collisional de-excitation, the molecular column densities depend only upon three combinations of these four parameters: $\zeta_p({\rm H})/n_{\rm H}$, $\chi_{\rm UV}/n_{\rm H}$, and $A_{\rm V}({\rm tot})$.  This behavior occurs because all bimolecular reactions and collisional cooling processes have rates (per unit volume) $\propto n_{\rm H}^2$, while all protoprocesses (e.g.\ photoionization, photodissociation, photoelectric heating) have rates that scale as $\chi_{\rm UV} n_{\rm H}$ and all cosmic-ray induced processes (e.g.\ cosmic-ray ionization, cosmic-ray heating) have rates $\propto \zeta_p({\rm H})n_{\rm H}.$  In this regime, therefore, we only needed to explore a three-dimensional parameter space consisting of $\chi_{\rm UV}/n_{50}$, $\zeta_p({\rm H})/n_{50}$, and $A_{\rm V}({\rm tot}),$ where $n_{\rm H} = 50\,n_{50}\,\rm{ cm}^{-2}$.  

With the exception mentioned in \S 2.5 below, all the calculations presented here were carried out for $n_{50}=1$, a typical value in the cold neutral medium.  Our standard grid comprises a total of 784 diffuse clouds models run for 8 values of $\chi_{\rm UV}/n_{50}$ in the range 0.2 to 10, 7 values of $\zeta_p({\rm H})/n_{50}$ in the range $6 \times 10^{-18} \, \rm s^{-1},$ to 6 $\times 10^{-15} \, \rm s^{-1},$  and 14 values of $A_{\rm V}({\rm tot})$ in the range $3 \times 10^{-4}$ to 3.

\re{Moreover, all the results presented below were obtained for ``steady-state" models, computed under the assumption that the cloud parameters vary over timescales that are much longer than that needed to reach chemical equilibrium.  Recent simulations of diffuse clouds, presented by Valdivia et al.\ (2016), suggest that the H$_2$ fraction may show departures from equilibrium in low-density regions.  Because the ion-neutral reactions leading to OH$^+$, $\rm H_2O^+$ and  ArH$^+$ are much faster than the grain-surface reactions that produce H$_2$, we expect these molecular ions to show abundances appropriate to the H$_2$ fraction, even if the H$_2$ fraction has not itself reached equilibrium.}  

\subsection{Enhanced-metallicity models}

We constructed a second complete grid of models to examine the dependence of the results on the adopted heavy-element abundances.  Here, we considered the case where all the heavy-element and dust abundances are increased by a factor 2.  This set of models was motivated by the presence of abundance gradients within the Galactic disk, and the fact that {\it Herschel} observations of molecular ions can probe the inner Galaxy as well as the solar neighborhood.

\subsection{Low-density model applicable to the warm neutral medium}

We also ran an additional model for a much lower density applicable to the warm neutral medium (WNM), $n_{\rm H} = 0.3 \,\rm cm^{-3}$.  Here, we were motivated by a suggestion 
(M\"uller et al.\ 2015) that argonium within the WNM might be a significant contributor to the observed ArH$^+$ absorption lines.  However, the predicted ArH$^+$ abundances within the WNM were found to be negligible relative to the observed values.  In particular, the predicted ArH$^+$ abundance relative to atomic hydrogen was only $\sim 5 \times 10^{-12}$ \re{for 
$\zeta_p({\rm H}) = 2 \times 10^{-16}\, \rm s^{-1}$, $\chi_{UV} = 1$, and $A_{\rm V}({\rm tot}) = 3 \times 10^{-4}$}; this value lies $\sim 2$ orders-of-magnitude below the abundance typically measured in the ISM.  The very low ArH$^+$ abundance predicted for the WNM reflects the extremely low molecular fraction present in the warm neutral phase, $f({\rm H}_2) \sim 10^{-7}$, which means that Ar$^+$ ions produced by cosmic-ray ionization are much more likely to suffer recombination than to react further with H$_2$.

\section{Results}

Figures 1 and 2 show examples of the predicted cloud structure, for four of the 784 standard-metallicity models. For the models shown here, $A_{\rm V}({\rm tot}) = 2$~mag in all four cases; $\chi_{\rm UV}/n_{50}$ = 1 (left panels) or 5 (left panels); and $\zeta_p({\rm H})/n_{50} = 2 \times 10^{-16}\,\rm s^{-1}$ (top panels) or $2 \times 10^{-15}\,\rm s^{-1}$ (bottom panels).  Blue, magenta, black, green and brown curves show the abundances of ArH$^+$, HCl$^+$, H$_2$Cl$^+$, OH$^+$, and H$_2$O$^+$ as a function of depth into the cloud.  The depth is represented as a visual extinction beneath the cloud surface, $A_{\rm V}$, plotted on the horizontal axis on a logarithmic scale.  The rightmost edge of the plot, with log$_{10}\,A_{\rm V} = 0,$ therefore represents the cloud center.  The abundances of the five molecular ions, each computed relative to H nuclei, are shown on a logarithmic scale in Figure 1 and on a linear scale in Figure 2 (normalized here relative to the peak abundance.)  Dotted red lines show the depths at which the molecular fraction, $f({\rm H}_2)$, reaches various values.  The molecular fraction increases rapidly towards the cloud center as the H$_2$ photodissociation rate drops precipitously due to self-shielding.

The abundances of the five molecular ions reach their peak values at different depths.  As discussed in \S 1, ArH$^+$ shows its peak predicted abundance at the smallest depth, in a region where $f({\rm H}_2) \sim 10^{-3.5}$.  The HCl$^+$ abundance peaks at an intermediate depth, where $f({\rm H}_2) \sim 10^{-2}$, while OH$^+$ and H$_2$Cl$^+$ are typically most abundant where $f({\rm H}_2) \sim 10^{-1}.$  The OH$^+$ and H$_2$O$^+$ abundance profiles often show two local maxima, one occurring at or near the cloud center where $f({\rm H}_2)$ is largest. 

These behaviors have been noted and discussed extensively by Neufeld et al.\ (2010) and H12 in the case of the oxygen-bearing molecular ions, and by Neufeld et al.\ (2012) in the case of the chlorine-bearing ions.  In particular, the OH$^+$/H$_2$O$^+$ and HCl$^+$/H$_2$Cl$^+$ ratios are controlled by the relative importance of H abstraction reactions [e.g.\ $\rm OH^+(H_2,H)H_2O^+$] and dissociative recombination [e.g.\ $\rm OH^+(e,H)O$].  

As explained previously in H12, the two local maxima in the OH$^+$ and H$_2$O$^+$ abundances are the result of two separate production mechanisms.  At small $f({\rm H}_2)$, the oxygen chemistry is initiated by the cosmic-ray ionization of H, followed by the reaction sequence 
$\rm H^+(O,H)O^+(H_2,H)OH^+(H_2,H)H_2O^+$; at larger H$_2$, the cosmic-ray ionization of H$_2$ is the initiating reaction, and is followed by  
$\rm H_2^+(H_2,H)H_3^+(O,H_2)OH^+(H_2,H)H_2O^+$.
The results shown in Figure 1 and 2 suggest that simultaneous observations of multiple molecular ions can be used to probe the relative amounts of gas with different molecular fractions within the diffuse ISM, a possibility that will be investigated 
further in \S 4 below.

In Figure 3, we show the column density ratio, $N({\rm ArH}^+)/N({\rm H})$, for the entire grid of standard-metallicity models.  Here, contours of log$_{10}\,[N({\rm ArH}^+)/N_{\rm H}]$ are shown (blue curves) in the space of CRIR (expressed as log$_{10}\,[\zeta_p({\rm H})/n_{50}]$ on the horizontal axes) and total extinction through the cloud (expressed as log$_{10}\,A_{\rm V}{\rm (tot)}$ on the vertical axes).  Different panels show the results obtained for various values of the UV radiation field ($\chi_{\rm UV}/n_{50}$.)  Solid red lines show the column-averaged molecular fraction, $f^{N}({\rm H}_2)=2N({\rm H}_2)/N_{\rm H}$, the superscripted $N$ being introduced in this notation to indicate a ratio of column densities rather than number densities.  For small CRIRs, $f^{N}({\rm H}_2)$ depends only on $A_{\rm V}{\rm (tot)}$ and is independent of $\zeta_p({\rm H})/n_{50}$, resulting in horizontal red contours for $f^{N}({\rm H}_2)$.  
At the highest CRIRs, however, the destruction of H$_2$ by cosmic-rays becomes important, the column-averaged molecular fraction starts to drop with increasing $\zeta_p({\rm H})/n_{50}$, and the red contours of   accordingly curve upwards. Figure 3 indicates that the maximum $N({\rm ArH}^+)/N({\rm H})$ attained at a given $\zeta_p({\rm H})/n_{50}$ occurs near $f^{N}({\rm H}_2) \sim 10^{-3.5}$ to $10^{-3}$.  \footnote{The results shown in Figure 3 can be compared directly with observations in cases where $N({\rm ArH}^+)$ has been obtained from submillimeter absorption observations and $N({\rm H})$ has been determined by observations of HI 21~cm absorption {\it combined with emission-line observations along nearby sight-lines.}  In cases where absorption-line observations of the HI 21~cm line are not accompanied by observations of emission, a direct comparison with observations is enabled by Figure 8 in Appendix A, which show contours of $N({\rm ArH}^+)/(N({\rm H})T_{\rm s2}^{-1}),$ where $100\,{T_{\rm s2}}$~K is the average HI spin temperature.}   \re{The $N({\rm ArH}^+)/N({\rm H})$ ratio drops significantly for $f^{N}({\rm H}_2)$ values above this range, but 
$N({\rm ArH}^+)/N({\rm H})$ ratio remains within a factor $\sim 2$ of its maximum value for $f^{N}({\rm H}_2)$ as small as $10^{-5}.$}   

In Figures 4 and 5, the $N({\rm ArH}^+)/N({\rm H})$ ratio is shown, once again, in the plane of log$_{10}\,[\zeta_p({\rm H})/n_{50}]$ and log$_{10}\,A_{\rm V}{\rm (tot)}$.  Now, contours of $N({\rm OH}^+)/N({\rm H})$ are also displayed, in green, along with contours of contours of $N({\rm H_2O}^+)/N({\rm H})$, in brown.  Figure 4 shows results for $\chi_{\rm UV}/n_{50}=1$, while separate panels in Figure 5 show results for $\chi_{\rm UV}/n_{50}=0.2$, 0.3, 2.0, and 3.0
In this plot, the contours are labeled in units of 10$^{-9}$, with bolder lines applying to the median values observed (\re{S14; Indriolo et al.\ 2015;} Gerin et al.\ 2016) within diffuse clouds in the Galactic disk (excluding the Galactic Center region): $N({\rm ArH}^+)/N({\rm H})= 3\times 10^{-10}$, $N({\rm OH}^+)/N({\rm H})= 1.2\times 10^{-8}$, $N({\rm H_2O}^+)/N({\rm H})= 2\times 10^{-9}.$

\section{Discussion}

The results shown in Figures 4 and 5 indicate that a single population of clouds cannot account for the typical $N({\rm ArH}^+)/N({\rm H})$, $N({\rm OH}^+)/N({\rm H})$ and $N({\rm H_2O}^+)/N({\rm H})$ abundance ratios observed in the diffuse ISM.  In particular, models with cloud parameters that can account for the OH$^+$ and H$_2$O$^+$ abundances underpredict the ArH$^+$ abundance by more than an order of magnitude.  For example, when $\chi_{\rm UV}/n_{50}=1$, the typical $N({\rm OH}^+)/N({\rm H})$ and $N({\rm H_2O}^+)/N({\rm H})$ ratios suggest a population of diffuse clouds of visual extinction $A_{\rm V}({\rm tot})\ \sim 0.15$~mag across each cloud, with a CRIR $\zeta_p({\rm H})/n_{50} \sim 3 \times 10^{-16}\rm s^{-1},$ within which the column-averaged molecular fraction $f^{N}({\rm H}_2) \sim 0.15$.  For the same total visual extinction, the typically-observed ArH$^+$ abundance would require a CRIR an order of magnitude larger, $\zeta_p({\rm H})/n_{50} \sim 3 \times 10^{-15}\rm s^{-1},$ while for the same CRIR required by the OH$^+$ and H$_2$O$^+$ abundance, the visual extinction would need to be $\simlt 0.005$~mag per cloud.  

These considerations suggest two possible explanations of the OH$^+$, H$_2$O$^+$ and ArH$^+$abundances observed in the diffuse ISM.   First, the CRIR could be a strongly-decreasing function of the visual extinction (i.e., the shielding column density), such that the regions very close to cloud surfaces, where ArH$^+$ is most abundant, are subject to a much larger cosmic-ray flux than the somewhat deeper regions where OH$^+$ and H$_2$O$^+$ are most abundant.   In this scenario, which is not addressed by our standard diffuse clouds models (in which the cosmic-ray flux is assumed to be constant throughout the cloud), 
enough ArH$^+$ might be produced in the surface layers of the same clouds responsible for the OH$^+$ and H$_2$O$^+$ absorption.  Second, the population of absorbing clouds within the Galactic disk might possess a distribution of sizes, with the ArH$^+$ absorption occurring primarily in the smaller-extinction clouds and the OH$^+$ and H$_2$O$^+$ absorption arising primarily in clouds of larger extinction. This second possibility may be favored by the observational fact (e.g. \re{S14, their Figure 6}; Neufeld et al.\ 2015, their Figure 11) that ArH$^+$ has a distribution (in line-of-sight velocity) that is different from that of any other molecular ion, whereas OH$^+$ and H$_2$O$^+$ show very similar absorption spectra. \

We have investigated the second scenario described above with a highly-idealized toy model in which the OH$^+$, H$_2$O$^+$ and ArH$^+$ absorption arises in a collection of clouds of two types: smaller clouds of visual extinction $A_{\rm V}{\rm (tot)_S}$, and larger clouds of visual extinction $A_{\rm V}{\rm (tot)_L}$.  These cloud types are assumed to account for a fraction $f_{\rm S}$ and $f_{\rm L} = 1 -f_{\rm S}$, respectively, of the HI mass, and both cloud types are assumed to be irradiated by the same cosmic-ray flux and UV radiation field.  Even if $\chi_{\rm UV}/n_{50}$ is specified, this model has four independent adjustable parameters -- $A_{\rm V}{\rm (tot)_S}$, $A_{\rm V}{\rm (tot)_L}$, $f_{\rm S}$, and $\zeta_p({\rm H})/n_{50}$ -- and is subject to three observational constraints:  $N({\rm ArH}^+)/N({\rm H})$, $N({\rm OH}^+)/N({\rm H})$ and $N({\rm H_2O}^+)/N({\rm H}).$  Accordingly, there is no unique solution for given $\chi_{\rm UV}/n_{50}$, but instead a set of solutions.  In Figure 4, 10 pairs of vertically-separated circles indicate ten possible solutions; one member of each pair represents a larger cloud type with log$_{10}\,A_{\rm V}{\rm (tot)} \ge -0.7$, and one member represents a smaller cloud type with log$_{10}\,A_{\rm V}{\rm (tot)} \le -1.7.$  The radius of each circle is proportional to the required mass-fraction within each cloud type.  By assumption, both cloud types are exposed to the same CRIR and thus each circle within a given pair has the same horizontal position.

An alternative presentation of the same information appears in Figure 6, where complete results are shown for multiple values of $\chi_{\rm UV}/n_{50}.$  Here, the following quantities are shown as a function of the assumed visual extinction across an individual smaller cloud: the required mass fraction of HI in the smaller clouds (top left), the column- averaged molecular fraction in the smaller (diamonds) and larger (asterisks) clouds (top right), the visual extinction required across an individual larger cloud (bottom left), and the required CRIR (bottom right).  Different colors represent different adopted values for $\chi_{\rm UV}/n_{50}.$  If we now focus on canonical values for the UV ISRF and gas density in the cold neutral medium, $\chi_{\rm UV} = n_{50}=1$ (black curve), we find that the typical abundances observed for OH$^+$, H$_2$O$^+$ and ArH$^+$ can be explained by a combination of two cloud types:  (1) smaller diffuse clouds, accounting for $\re{40 - 75}\%$ of the gas mass and having a visual extinction $\le 0.02$~mag and a column-averaged molecular fraction in the range $3 \times 10^{-5}$ to $10^{-2}$; and (2) larger diffuse clouds, accounting for $\re{25 - 60}\%$ of the gas mass and having a visual extinction $\ge \re{0.2}$~mag and a column-averaged molecular fraction $\sim 0.2$.  \re{In the case of the smaller diffuse clouds, acceptable fits can be obtained for the smallest values of $A_{\rm V}{\rm (tot)}$ that we considered (and thus the smallest values of $f^{N}({\rm H}_2)$ predicted in any of our models.)  The observed ArH$^+$ abundances are therefore entirely consistent with UV observations of H and H$_2$ in diffuse clouds; such observations indicate that most clouds have $f^{N}({\rm H}_2)$ either greater than $10^{-2}$ or smaller than $10^{-4},$ and that relatively few clouds have $f^{N}({\rm H}_2)$ in the intermediate range ($10^{-2} - 10^{-4}$) where H$_2$ self-shielding results in a very strong dependence of $f^{N}({\rm H}_2)$ upon $A_{\rm V}{\rm (tot)}$ (e.g. Liszt 2015; and references therein).}   

The highly-idealized two-component model presented here requires a primary CRIR in the range $\zeta_p({\rm H}) = \re{4 - 8} \times 10^{-16} \, \rm s^{-1}$, \re{for $\chi_{\rm UV} = n_{50}=1$}  (For the enhanced-metallicity models, \re{the $A_{\rm V}{\rm (tot)}$ values required for the larger clouds are somewhat smaller, 
but the required ionization rates are very similar.})
\re{This range of primary CRIR lies slightly above} the value $\sim 3 \times 10^{-16} \, \rm s^{-1}$ needed to account for $N({\rm OH}^+)/N({\rm H})$ and $N({\rm H_2O}^+)/N({\rm H})$  with a single-component model, because in the two-component model, only \re{25 - 60}$\%$ of the HI is in the larger clouds that contribute most of the OH$^+$ and $\rm H_2O^+$.  The CRIR required for the two-component model is also a factor $\re{2.5-5}$ larger than the average value inferred by Indriolo \& McCall (2012) from observations of H$_3^+$ in clouds of somewhat larger size than those considered here.  This discrepancy 
 may reflect the idealized nature of our two-component model or uncertainties in the chemistry; alternatively, it may suggest that shielding does modulate the cosmic-ray flux in the regions where H$_3^+$ is present.  The question of the CRIR and its depth-dependence will be investigated further in a future paper.

\section{Summary}

\noindent 1.  We have presented a general parameter study, in which the argonium abundance is predicted using a model for the physics and chemistry of diffuse interstellar gas clouds.
Results have been obtained as a function of UV radiation field, $\chi_{\rm UV}/n_{50}$, in the range 0.2 to 10, cosmic-ray ionization rates, $\zeta_p({\rm H})/n_{50},$ in the range $6 \times 10^{-18}\, \rm s^{-1}$ to 6 $\times 10^{-15}\, \rm s^{-1},$ and cloud extinctions, $A_{\rm V}{\rm (tot)}$, in the range $3 \times 10^{-4}$ to 3.
\vskip 0.1 true in
\noindent 2.  No single set of cloud parameters provides an acceptable fit to the typical ArH$^+$, OH$^+$ and $\rm H_2O^+$ abundances observed in diffuse clouds within the Galactic disk.  Instead, the observed abundances suggest that ArH$^+$ resides primarily in separate population of small clouds of $A_{\rm V}{\rm (tot)} \le 0.02$~mag, within which the column-averaged molecular fraction is at \re{most} $10^{-2}$, while OH$^+$ and $\rm H_2O^+$ reside primarily in somewhat larger clouds with a column-averaged molecular fraction $\sim 0.2$.  This analysis confirms our previous suggestion (S14) that the argonium molecular ion is a unique tracer of almost purely {\it atomic} gas.
\vskip 0.1 true in
\noindent 3.  Our simultaneous fit to the observed abundances of ArH$^+$, OH$^+$ and $\rm H_2O^+$ suggests a primary CRIR in the range $\zeta_p({\rm H}) = 4 - \re{8} \times 10^{-16} \, \rm s^{-1}$ for assumed metallicities between 1 and 2 times that typical of the solar neighborhood.

\begin{acknowledgements}
We gratefully acknowledge the support of a grant from NASA's Astrophysical Data Analysis Program (ADAP). \re{We thank M.~Gerin and the anonymous referee for several valuable comments about an earlier version the manuscript.}

\end{acknowledgements}

\begin{figure}
\includegraphics[width=15 cm]{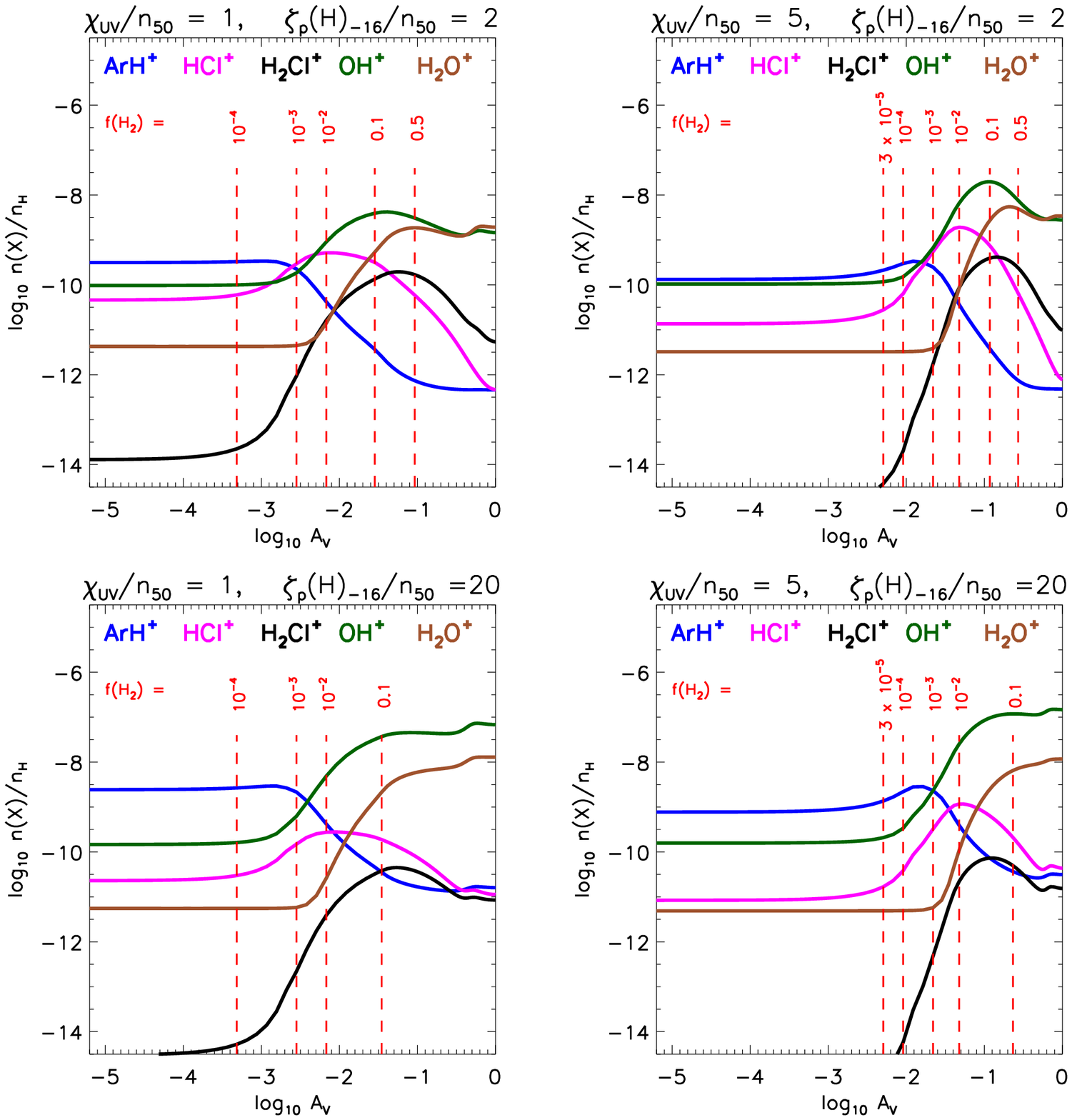}
\caption{Examples of the predicted cloud structure, for four of the 784 standard-metallicity models.  {\it Left panels}: $\chi_{\rm UV}/n_{50} = 1$; {\it right panels}: $\chi_{\rm UV}/n_{50} = 5$; {\it top panels}: $\zeta_p({\rm H})/n_{50} = 2 \times 10^{-16}\,\rm s^{-1}$; {\it bottom panels}: $\zeta_p({\rm H})/n_{50}=2 \times 10^{-15}\,\rm s^{-1}$.  Blue, magenta, black, green and brown curves show the abundances of ArH$^+$, HCl$^+$, H$_2$Cl$^+$, OH$^+$, and H$_2$O$^+$ as a function of depth into the cloud, the depth being represented as a visual extinction beneath the cloud surface, $A_{\rm V}$, and plotted on the horizontal axis on a logarithmic scale.  The total visual extinction across the cloud is $A_{\rm V}({\rm tot}) = 2$~mag in all four cases. The abundances of the five molecular ions, each computed relative to H nuclei, are shown on a logarithmic scale.}
\end{figure}

\begin{figure}
\includegraphics[width=15 cm]{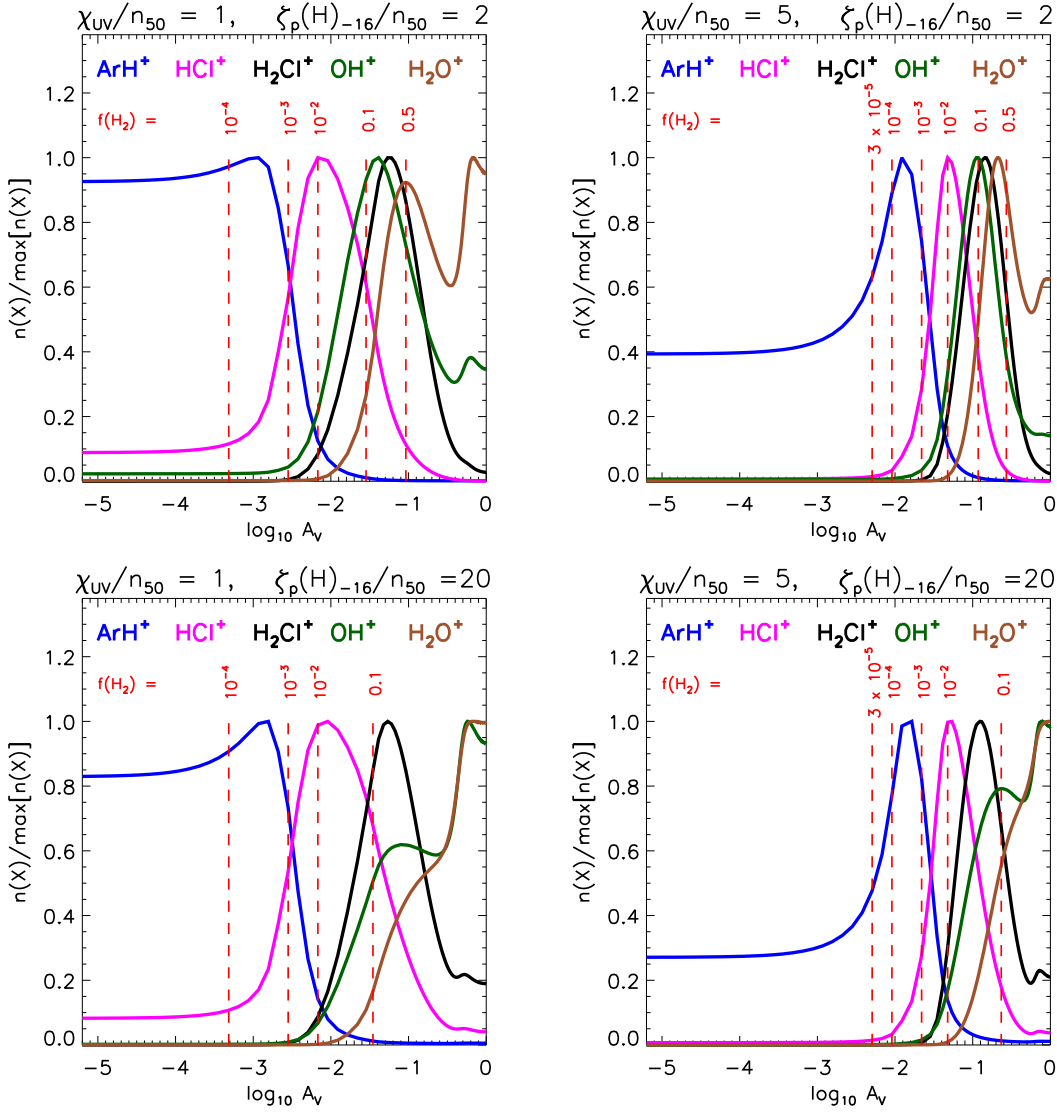}
\caption{Same as Figure 1, but with the abundances shown on a linear scale and normalized with respect to the peak abundance.}
\end{figure}

\begin{figure}
\includegraphics[width=13 cm]{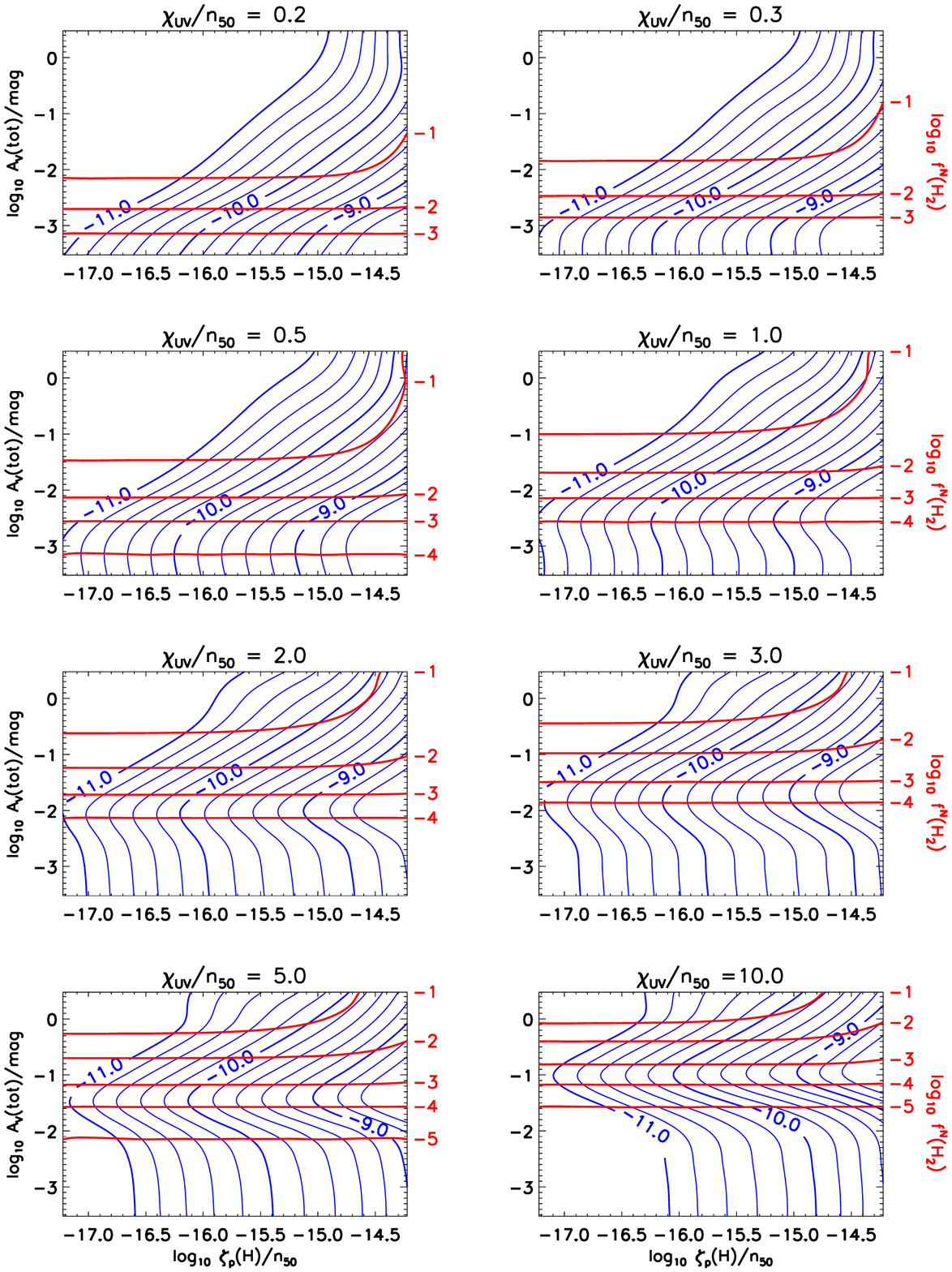}
\caption{Column density ratio, $N({\rm ArH}^+)/N({\rm H})$, for the entire grid of standard-metallicity models.  Contours, labeled by log$_{10}\,[N({\rm ArH}^+)/N_{\rm H}],$ are shown (blue curves) in the space of CRIR (expressed as log$_{10}\,[\zeta_p({\rm H})/n_{50}]$ on the horizontal axes) and total extinction through the cloud (expressed as log$_{10}\,A_{\rm V}{\rm (tot)}$ on the vertical axes).  Different panels show the results obtained for various values of the UV radiation field ($\chi_{\rm UV}/n_{50}$.)  Solid red lines show the column-averaged molecular fraction, $f^{N}({\rm H}_2)=2N({\rm H}_2)/N_{\rm H}$.}
\end{figure}

\begin{figure}
\includegraphics[width=13 cm]{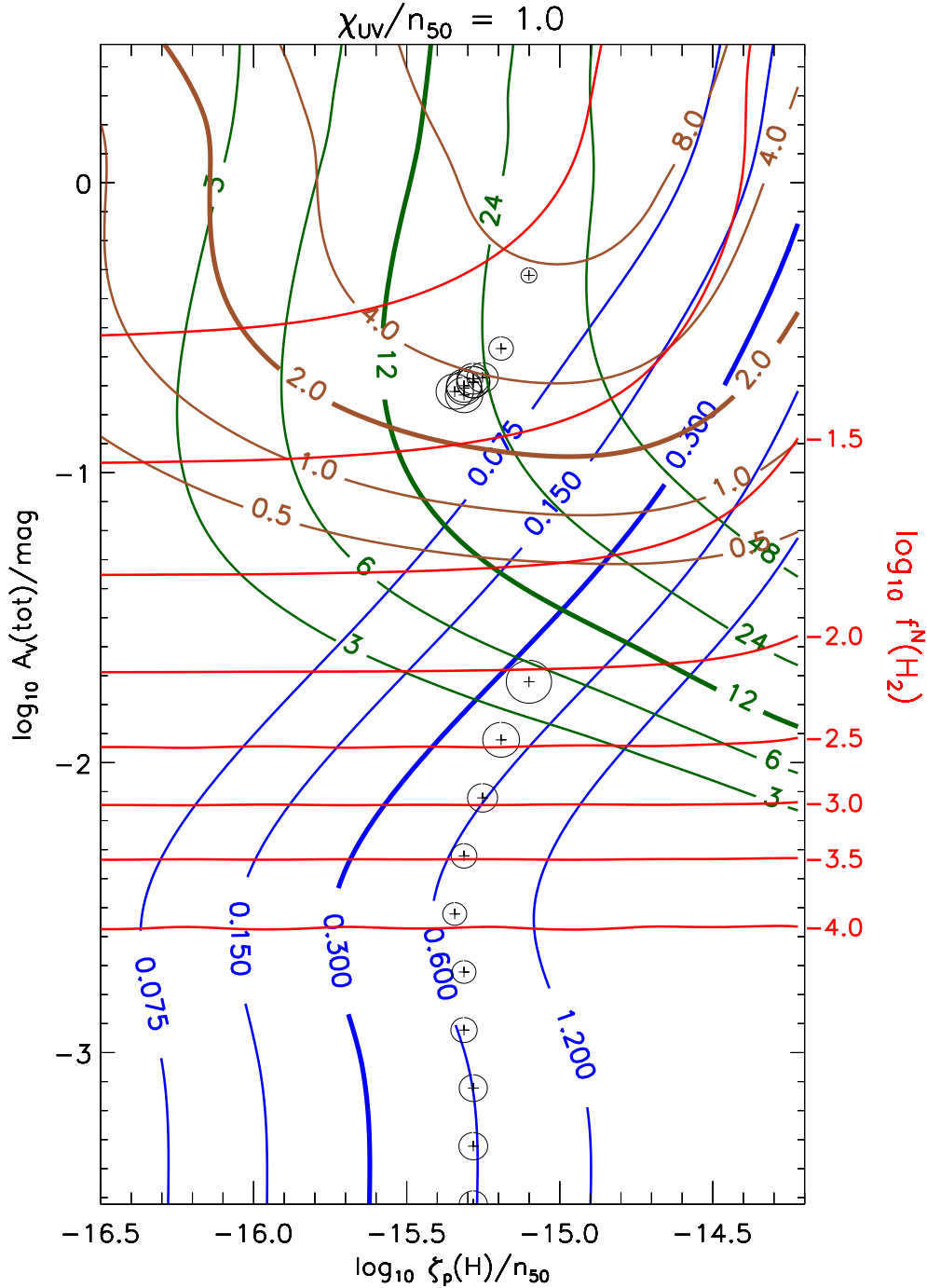}
\caption{Contours of the $N({\rm ArH}^+)/N({\rm H})$, $N({\rm OH}^+)/N({\rm H})$, and $N({\rm H_2O}^+)/N({\rm H})$ abundance ratios (blue, green and brown curves, respectively) in the plane of log$_{10}\,[\zeta_p({\rm H})/n_{50}]$ and log$_{10}\,A_{\rm V}{\rm (tot)}$.  Contours are labeled in units of 10$^{-9}$, with bolder lines applying to the median values observed within diffuse clouds in the Galactic disk (Gerin et al.\ 2016).  Results are shown for $\chi_{\rm UV}/n_{50}=1.$  Black circles indicate best-fit parameters for the two-component model described in the text.}
\end{figure}

\begin{figure}
\includegraphics[width=15 cm]{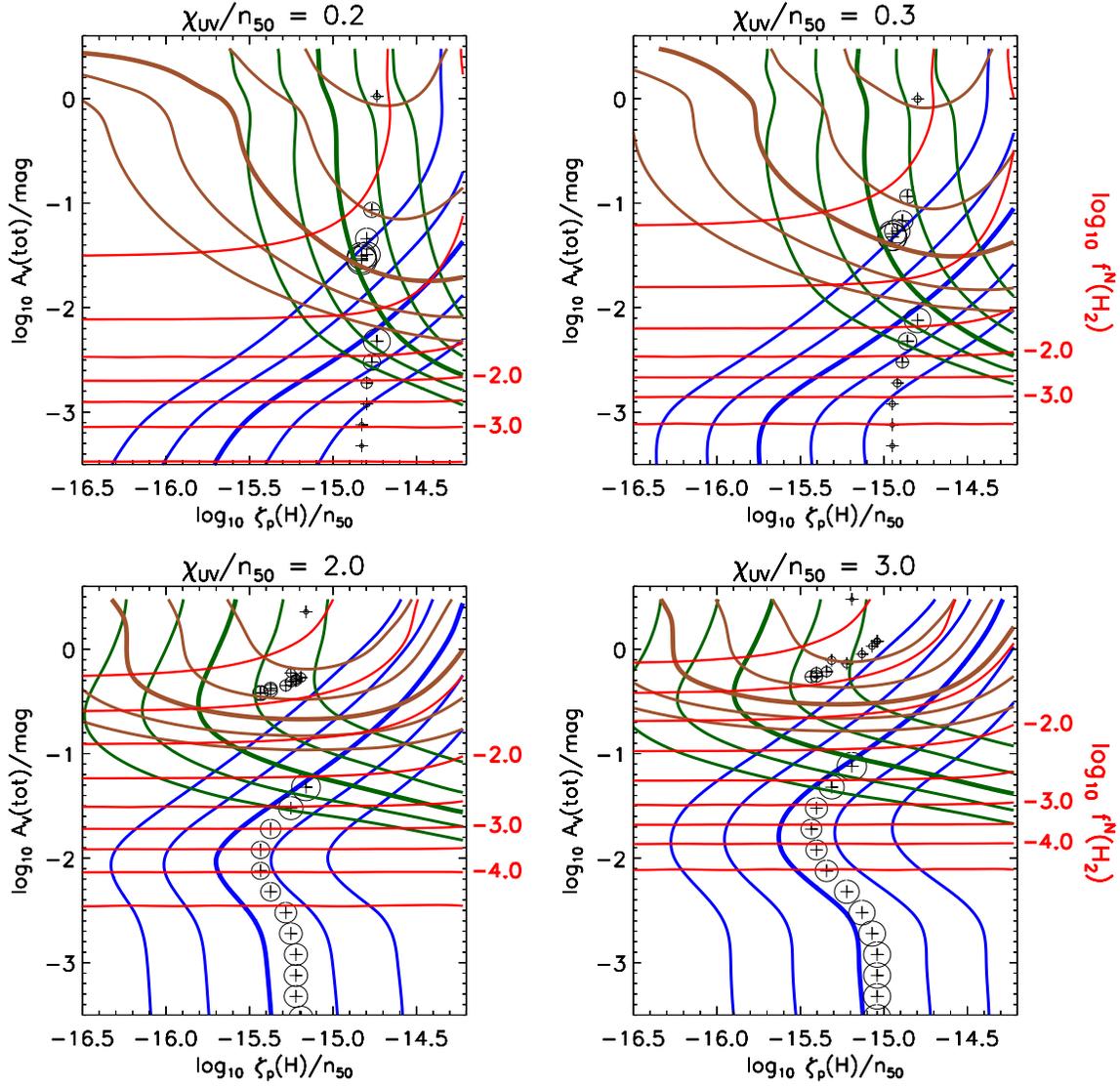}
\caption{Same as Figure 4, but for $\chi_{\rm UV}/n_{50}=0.2$, 0.3, 2.0, and 3.0.  For clarity, the blue, green and brown contours are unlabeled here, but the values are identical to those appearing in Figure 4 and increase from left to right.}
\end{figure}

\begin{figure}
\includegraphics[width=15 cm]{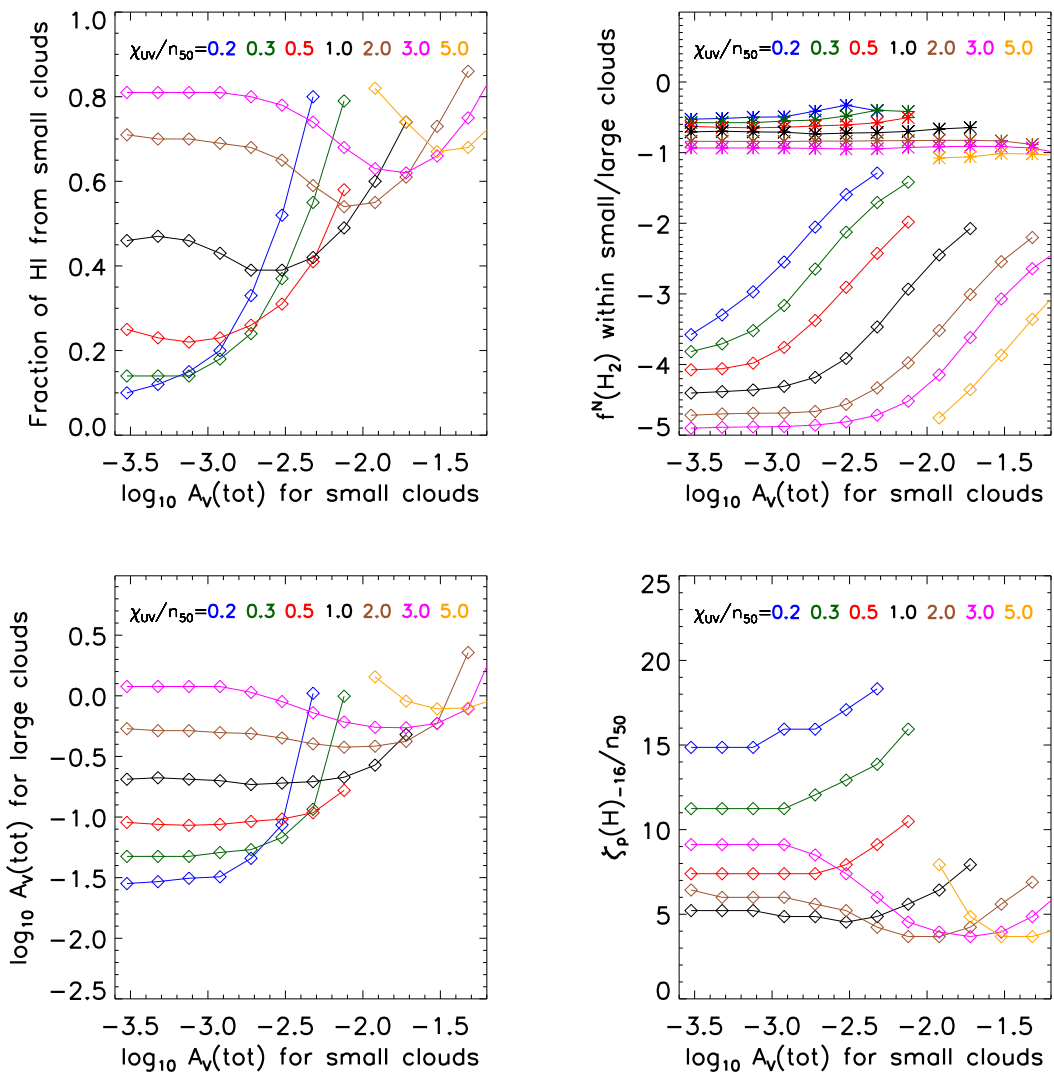}
\caption{Best-fit cloud parameters in the two-component model (see text), as a function of the assumed visual extinction across an individual smaller cloud. {\it Top left}: mass fraction of HI in the smaller clouds; {\it top right}: column-averaged molecular fraction in the smaller (diamonds) and larger (asterisks) clouds; {\it bottom left}: visual extinction across an individual larger cloud; {\it bottom right}: cosmic-ray ionization rate.  Different colors represent different adopted values for $\chi_{\rm UV}/n_{50}.$  }
\end{figure}

\clearpage

\begin{appendix}

\section{The HI 21~cm spin temperature}
In cases where an HI column density is to be determined from absorption-line observations of the HI 21~cm line alone, some assumption must be made about the average 
spin temperature along the sight-line, $T_{\rm s}= 100\,T_{\rm s2}$~K. 
The inferred HI column density is proportional to the average spin temperature, the 
appropriate average being a HI-density-weighted geometric mean:
$$ T_{\rm s} = N({\rm H}) / \int n({\rm H}) T^{-1} ds.$$
Here, $ds$ is the element of length along the line-of-sight, and the local spin temperature is assumed equal to the gas kinetic temperature, $T$.

In Figure 7, we present predictions for the average spin temperature, $T_{\rm s}$, as a function of $\chi_{\rm UV}/n_{50}$ and log$_{10}\,[\zeta_p({\rm H})/n_{50}].$  Red, blue and green curves apply to $A_{\rm V}{\rm (tot)}$ = 0.01, 0.1 and 1, respectively.
Figure 8 shows contours of $N({\rm ArH}^+)/(N({\rm H})T_{\rm s2}^{-1})$
in manner analogous to Figure 3.

\begin{figure}
\includegraphics[width=14 cm]{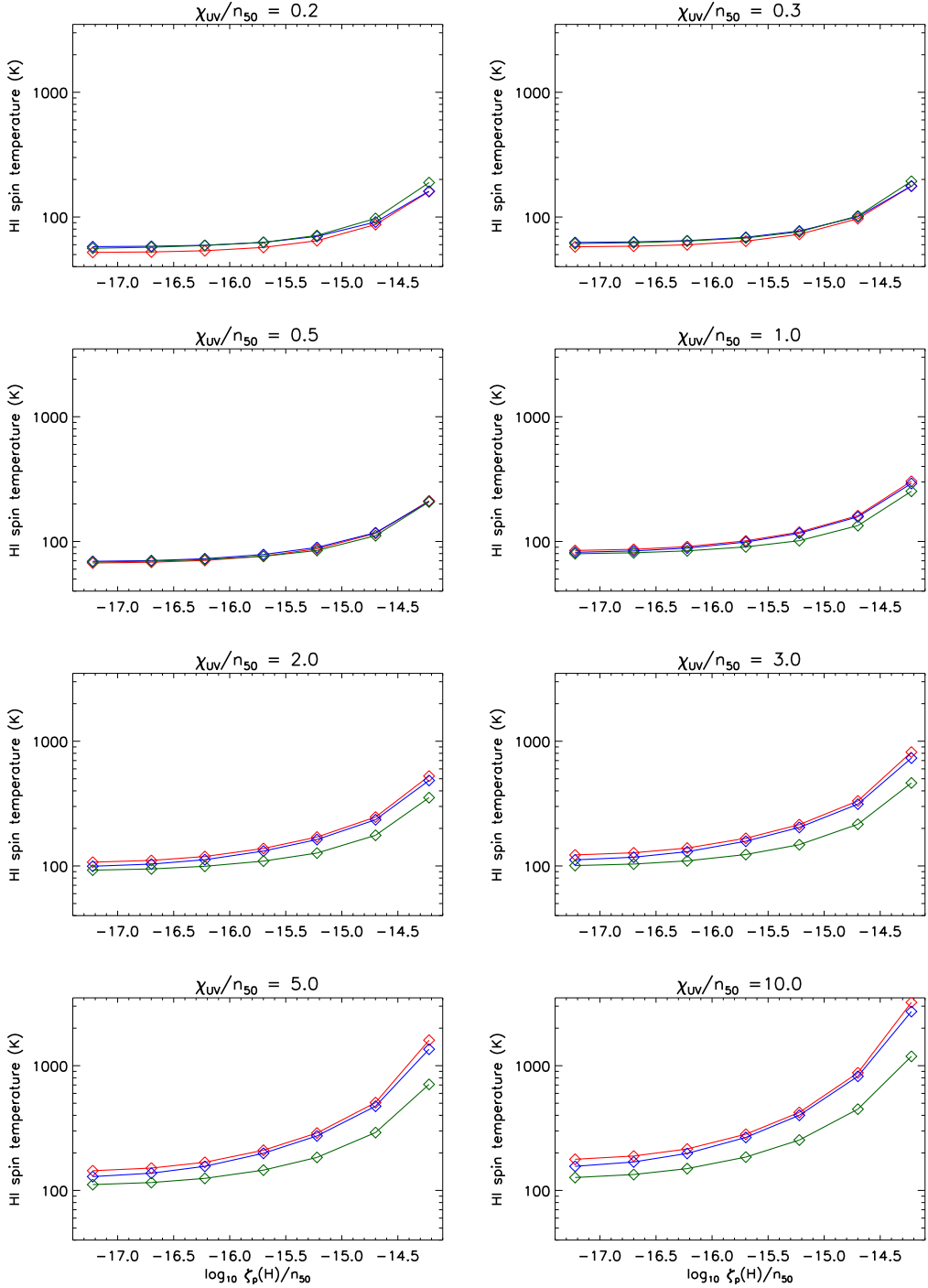}
\caption{Line-of-sight average spin temperature, $T_{\rm s}$, for 
standard-metallicity models, as a function of $\chi_{\rm UV}/n_{50}$ and for several values of log$_{10}\,[\zeta_p({\rm H})/n_{50}].$  Red, blue and green curves apply to $A_{\rm V}{\rm (tot)}$ = 0.01, 0.1 and 1, respectively.}
\end{figure}

\begin{figure}
\includegraphics[width=13 cm]{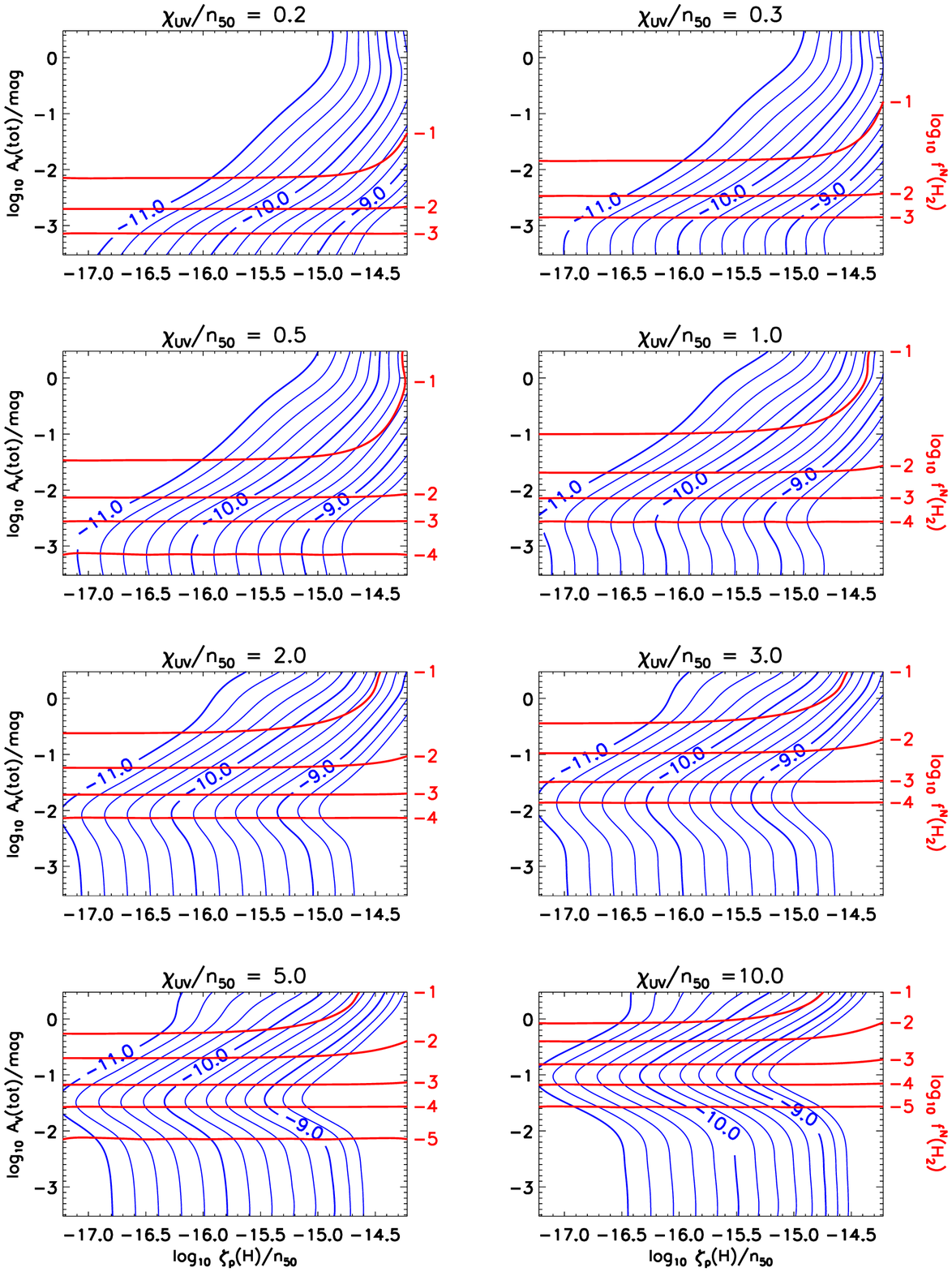}
\caption{Column density ratio, $N({\rm ArH}^+)/[N({\rm H})T_{\rm s2}^{-1}]$, for the entire grid of standard-metallicity models.  Contours, labeled by log$_{10}\,[N({\rm ArH}^+)/(N({\rm H})T_{\rm s2}^{-1})],$ are shown (blue curves) in the space of CRIR (expressed as log$_{10}\,[\zeta_p({\rm H})/n_{50}]$ on the horizontal axes) and total extinction through the cloud (expressed as log$_{10}\,A_{\rm V}{\rm (tot)}$ on the vertical axes).  Different panels show the results obtained for various values of the UV radiation field ($\chi_{\rm UV}/n_{50}$.)  Solid red lines show the column-averaged molecular fraction, $f^{N}({\rm H}_2)=2N({\rm H}_2)/N_{\rm H}$.}
\end{figure}
\end{appendix}


\begin{thebibliography}{}

\bibitem[Abrahamsson et al.(2007)]{2007ApJ...654.1171A} Abrahamsson, E.,
Krems, R.~V., \& Dalgarno, A.\ 2007, \apj, 654, 1171

\bibitem[Alekseyev et al.(2007)]{2007PCCP....9.5088A} Alekseyev, A.~B., Liebermann, H.-P., \& Buenker, R.~J.\ 2007, Physical Chemistry Chemical Physics (Incorporating Faraday Transactions), 9, 5088 

\bibitem[Barinovs et al.(2005)]{2005ApJ...620..537B} Barinovs, {\u G}., van
Hemert, M.~C., Krems, R., \& Dalgarno, A.\ 2005, \apj, 620, 537

\bibitem[Barlow et al.(2013)]{2013Sci...342.1343B} Barlow, M.~J., Swinyard, B.~M., Owen, P.~J., et al.\ 2013, Science, 342, 1343 

\bibitem[Burton et al.(2015)]{2015ApJ...811...13B} Burton, M.~G., Ashley,
M.~C.~B., Braiding, C., et al.\ 2015, \apj, 811, 13

\bibitem[Cartledge et al.(2004)]{2004ApJ...613.1037C} Cartledge, S.~I.~B., Lauroesch, J.~T., Meyer, D.~M., \& Sofia, U.~J.\ 2004, \apj, 613, 1037 

\bibitem[Dalgarno et al.(1999)]{1999ApJS..125..237D} Dalgarno, A., Yan, M., \& Liu, W.\ 1999, \apjs, 125, 237 

\bibitem[de Ruette et al.(2016)]{2016ApJ...816...31D} de Ruette, N.,
Miller, K.~A., O'Connor, A.~P., et al.\ 2016, \apj, 816, 31

\bibitem[Draine(1978)]{1978ApJS...36..595D} Draine, B.~T.\ 1978, \apjs, 36, 595 

\bibitem[Gerin et al.(2015)]{2015A&A...573A..30G} \re{Gerin, M., Ruaud, M., Goicoechea, J.~R., et al.\ 2015, \aap, 573, A30} 

\bibitem[Gerin et al.(2016)]{2016arXiv160102985G} Gerin, M., Neufeld, D.~A., \& Goicoechea, J.~R.\ 2016, \araa, in press (arXiv:1601.02985) 

\bibitem[Glassgold \& Langer(1973)]{1973ApJ...186..859G} Glassgold, A.~E., \& Langer, W.~D.\ 1973, \apj, 186, 859 

\bibitem[He et al.(2015)]{2015ApJ...801..120H} He, J., Shi, J., Hopkins,
T., Vidali, G., \& Kaufman, M.~J.\ 2015, \apj, 801, 120

\bibitem[Hollenbach et al.(2012)]{2012ApJ...754..105H} Hollenbach, D., Kaufman, M.~J., Neufeld, D., Wolfire, M., \& Goicoechea, J.~R.\ 2012, \apj, 754, 105 (H12)

\bibitem[Indriolo \& McCall(2012)]{2012ApJ...745...91I} Indriolo, N., \& McCall, B.~J.\ 2012, \apj, 745, 91 

\re{\bibitem[Indriolo et al.(2015)]{2015ApJ...800...40I} Indriolo, N., Neufeld, D.~A., Gerin, M., et al.\ 2015, \apj, 800, 40}

\bibitem[Liszt(2015)]{2015ApJ...799...66L} \re{Liszt, H.~S.\ 2015, \apj, 799, 66} 

\bibitem[McElroy et
al.(2013)]{2013A&A...550A..36M} McElroy, D., Walsh, C., Markwick, A.~J., et al.\ 2013, \aap, 550, A36

\bibitem[Mitchell et al.(2005)]{2005JPhB...38L.175M} Mitchell, J.~B.~A., Novotny, O., LeGarrec, J.~L., et al.\ 2005, Journal of Physics B Atomic Molecular Physics, 38, L175 

\bibitem[M{\"u}ller et al.(2013)]{2013AIPC.1545...96M} M{\"u}ller, H.~S.~P., Endres, C.~P., Stutzki, J., \& Schlemmer, S.\ 2013, American Institute of Physics Conference Series, 1545, 96 

\bibitem[M{\"u}ller et al.(2015)]{2015A&A...582L...4M} M{\"u}ller, H.~S.~P., Muller, S., Schilke, P., et al.\ 2015, \aap, 582, L4 

\bibitem[Neufeld et 
al.(2010)]{2010A&A...521L..10N} Neufeld, D.~A., Goicoechea, J.~R., Sonnentrucker, P., et al.\ 2010, \aap, 521, L10 

\bibitem[Neufeld et al.(2012)]{2012ApJ...748...37N} Neufeld, D.~A., Roueff, E., Snell, R.~L., et al.\ 2012, \apj, 748, 37 


\bibitem[Neufeld et al.(2015)]{2015A&A...577A..49N} \re{ Neufeld, D.~A., Godard, B., Gerin, M., et al.\ 2015, \aap, 577, A49}

\bibitem[Novotn{\'y} et al.(2013)]{2013ApJ...777...54N} Novotn{\'y}, O.,
Becker, A., Buhr, H., et al.\ 2013, \apj, 777, 54

\bibitem[Roueff et al.(2014)]{2014A&A...566A..30R} Roueff, E., Alekseyev, A.~B., \& Le Bourlot, J.\ 2014, \aap, 566, A30 

\bibitem[Schilke et al.(2014)]{2014A&A...566A..29S} Schilke, P., Neufeld, D.~A., M{\"u}ller, H.~S.~P., et al.\ 2014, \aap, 566, A29 (S14)

\bibitem[Sofia et al.(2004)]{2004ApJ...605..272S} Sofia, U.~J., Lauroesch, J.~T., Meyer, D.~M., \& Cartledge, S.~I.~B.\ 2004, \apj, 605, 272 

\bibitem[Valdivia et al.(2016)]{2016A&A...587A..76V} \re{Valdivia, V., Hennebelle, P., G{\'e}rin, M., \& Lesaffre, P.\ 2016, \aap, 587, A76}


\bibitem[Wiesenfeld \& Goldsmith(2014)]{2014ApJ...780..183W} Wiesenfeld, L., \& Goldsmith, P.~F.\ 2014, \apj, 780, 183

\bibitem[Wolfire et al.(2010)]{2010ApJ...716.1191W} Wolfire, M.~G.,
Hollenbach, D., \& McKee, C.~F.\ 2010, \apj, 716, 1191

\end{thebibliography}
\end{document}